\documentclass[12pt,preprint]{aastex}
\usepackage{graphicx}
\usepackage{amsmath}
\newcommand{\Msun}{~M_\odot}

\newcommand{\cmc}{\rm ~cm^{-3}}
\newcommand{\kms}{\rm ~km~s^{-1}}

\begin{document}
\title{GAMMA-RAY EMISSION FROM SUPERNOVA REMNANT INTERACTION WITH MOLECULAR CLUMPS}
\author{Xiaping Tang and Roger A. Chevalier}
\affil{Department of Astronomy, University of Virginia, P.O. Box 400325, \\
Charlottesville, VA 22904-4325; xt5ur@virginia.edu, rac5x@virginia.edu}

\begin{abstract}
Observations of the middle-aged supernova remnants IC 443, W28 and W51C indicate that
the brightnesses at GeV and TeV energies are correlated with each other and with regions of molecular clump interaction, but not with the radio synchrotron brightness.
We suggest that the radio emission is primarily associated with a radiative shell in the
interclump medium of a molecular cloud, while the $\gamma$-ray emission is primarily
associated with the interaction of the radiative shell with molecular clumps.
The shell interaction produces a high pressure region, so that the $\gamma$-ray luminosity
can be approximately reproduced even if shock acceleration of particles is not efficient,
provided that energetic particles are trapped in the cooling region.
In this model, the spectral shape $\ga 2$ GeV is determined by the spectrum of cosmic
ray protons. Models in which diffusive shock acceleration determines the spectrum tend to underproduce TeV emission because of the limiting particle energy that is attained.  

\end{abstract}


\section{INTRODUCTION}

A highlight of high energy $\gamma$-ray astronomy, involving space-based observations at
GeV energies and ground-based observations at TeV energies, has been the detection of
middle-aged supernova remnants (SNRs) interacting with molecular clouds (MCs) \citep[see][for reviews]{uchiyama11,fernandez13}.
Following pioneering observations with {\it EGRET} \citep{esposito96},
the {\it Fermi} and {\it AGILE} observatories observed $\rm GeV$ $\gamma$-ray emission from several middle-aged SNRs which are interacting with MCs, including W51C \citep{Abdo09}, W44 \citep{Abdo10a},  IC 443 \citep{tavani10,Abdo10b}, and W28 \citep{Abdo10c}. Of these, TeV emission is also detected in W51C \citep{Aleksic12}, IC 443 \citep{Albert07} and  W28 \citep{Aharonian08}.
In the cases of W28 \citep{Aharonian08} and W44 \citep{uchiyama12}, there is $\gamma$-ray emission external to the remnants that may be associated with the remnants.
The high energy emission from these sources has generally been interpreted in terms
of pion-decays from cosmic ray (CR) interactions.

Two scenarios have been proposed to explain the properties of these middle-aged SNRs  associated with MC interaction. 
In one, relativistic particles escape from a SNR and interact with a nearby MC; TeV emission is produced by interaction between escaping CR particles and the MC, while GeV emission is produced by interaction between galactic CR background particles and the MC \citep{Gabici09,torres10}. 
In view of the two components, this scenario naturally produces double peaked $\gamma$-ray spectra.

The other scenario, discussed here, involves radiative shock waves
\citep{Chevalier77,B&C82,Chevalier99,bykov00,Uchiyama10}.
The compressed region downstream from a radiative shock is promising because of the high particle number density and energy density.
\cite{Uchiyama10} presented a crushed cloud model for W44, IC 443, and W51C, based on remnant parameters from \cite{reach05} for W44.
In this view, the SNR has a $500\kms$ nonradiative shock in most of the volume and drives 
 $100\kms$ radiative shock waves into clouds with a density of $200$ cm$^{-3}$.
Ambient CRs experience diffusive shock acceleration (DSA) as well as compression in the shell.
The cooling region downstream from the shock front is presumed to be the site of $\gamma$-ray emission and radio synchrotron emission.

Here, we examine the $\gamma$-ray emission properties of IC 443, W28, and W51C in order to
gain insight into the emission processes (Section~2).
 In Section~3, we discuss the structure of the magnetically supported radiative shell and the interaction region between shell and the MC clump \citep{Chevalier99}.
In Section~4, we model the evolution of relativistic protons in both  regions and  calculate their $\pi^0$-decay emission. We compare our results with the observations of  all three remnants and discuss the results in Section 5. 

\section{EMISSION PROPERTIES}

Three well observed remnants, IC 443, W51C and W28, have been detected at both GeV and TeV energies.  The following points can be made based on these objects:

1)  The GeV and TeV emission regions are well correlated with each other.
In the case of IC 443, the centroids of the GeV and TeV emission differ, but the spatially extended regions of emission largely overlap in the southeast part of the remnant  \citep{Abdo10b}.
For W28, Figure 1 of \cite{Aharonian08} shows that the H.E.S.S. TeV source J1801-233 is correlated with the GeV source
from {\it EGRET} and both are in the eastern part of the remnant.
For W51C, the TeV emission measured by {\it MAGIC} correlates with the GeV emission observed by {\it Fermi}, although the situation is complicated by a possible pulsar wind nebula \citep[Figure 4 of][]{Aleksic12}.

2)  The high energy $\gamma$-ray emission is spatially correlated with regions of molecular shock interaction.  
In IC 443, the GeV and TeV emission are correlated with the region where there are shocked
CO clumps \citep[Figure 5 of][]{Abdo10b}.
In W28, the H.E.S.S J1801-233 source closely overlaps a region of shocked molecular
emission \citep[Figures 1 and 2 of][]{nicholas12}.
Figure 1 of \cite{uchiyama11} shows that the highest surface brightness GeV emission in W51C correlates
well with a region of shocked MC \citep{koo97}.

3)  The three remnants have similar $\gamma$-ray spectra from GeV to TeV energies and are distinct from younger remnants \citep[Figure 1 in][]{cardillo12}.  The spectra  do not clearly show evidence for more than one component.  The shape of the spectra at low energies is consistent with $\gamma$-rays from $\pi^0$-decays \citep{Ackermann13}.
The similar GeV to TeV flux ratios for the remnants can also be seen in
Figure 1 of \cite{fernandez13}. 

4)    The brightest $\gamma$-ray emission is not well correlated with nonthermal radio emission, e.g., IC 443 and W28 \citep[Figure 1 of][]{uchiyama11}, although
\cite{uchiyama11} notes that the spatial extent of the GeV emission is comparable to
the radio remnant for these 3 remnants.
However,  the radio structure in IC 443 is well correlated with optical emission from radiative shock fronts with velocities of $65 - 100 \kms$ and preshock densities of $10 - 20$ cm$^{-3}$ \citep{fesen80}.  
In W28, the radio emission is brightest on the north side of the remnant \citep{brogan06},
while the high energy emission is to the east and northeast.
In W51C, there is radio emission overlapping the GeV emission, but there is also more radio emission to the north \citep[Figure 1 of][]{uchiyama11}.

The close correlation of GeV and TeV emission with MC interactions
as well as the single component spectra suggests that
the emission is not from escaping CRs for these 3 remnants, but from radiative shock
waves.
However, there is a distinction between regions of high radio brightness, which are not particularly correlated with shocked molecular emission, and regions of high $\gamma$-ray brightness, which are.
The emission can be interpreted in the context of the MC interaction
model of \cite{Chevalier99}.
The radio emission is from the radiative shell formed when the shock front moves into the interclump medium (ICM) of the MC with a  density of $\sim 5-25$ cm$^{-3}$. 
This shell may be the source of some high energy $\gamma$-ray emission, but the brightest emission is from the regions of molecular clump interaction. The collision of the radiative shell
with a molecular clump  produces a region of especially high energy density which is promising for $\gamma$-ray emission. 

\section{CLUMP INTERACTION}

We assume that
the radiative shell in the ICM is thin and supported by magnetic pressure. 
The shock wave is strong, so $B_{ts}^2/8\pi=\rho_0V_s^2$, where $B_{ts}$ is the tangential magnetic field in the shell,  $\rho_0=\xi_Hn_0m_p$ is the density of the ICM, and $V_s$ is the shock velocity. We assume a helium number abundance of $10\%$ hydrogen nuclei ($\xi_H=1.4$); from here on, ambient number density  refers to hydrogen nuclei.
Based on the conservation of mass and magnetic flux in the shell \citep{Chevalier77}, the density in the shell is
\begin{equation}
\rho_s=\frac{2}{3}\alpha\rho_0\frac{R_s^3-R_b^3}{R_s^3-R_b^2 R_s},
\end{equation}
where $\alpha=B_{ts}/B_{t_0}$ is the compression factor, $R_s$ is the shock radius, and $R_b$ is the radius at the cooling time $t_b=(t_{sf}+t_{PDS})/2$, where $t_{sf}$ and $t_{PDS}$ are as in \cite{Cioffi88}.
For an ambient magnetic field $B_0$ tangled on a scale much smaller than $R_s$, $B_{t_0}^2=(2/3)B_0^2$.
Zeeman measurements of diffuse and molecular clouds show that the total magnetic field $B$ within clouds tends to be constant $ \la 10~\mu$G ($n_0<300\cmc$) and $\propto n_0^\kappa$ where $\kappa\approx 0.65$
 ($n_0>300\cmc$) \citep{Crutcher12}.

\begin{figure}[htb]
 \begin{center}
  \includegraphics[width=\textwidth]{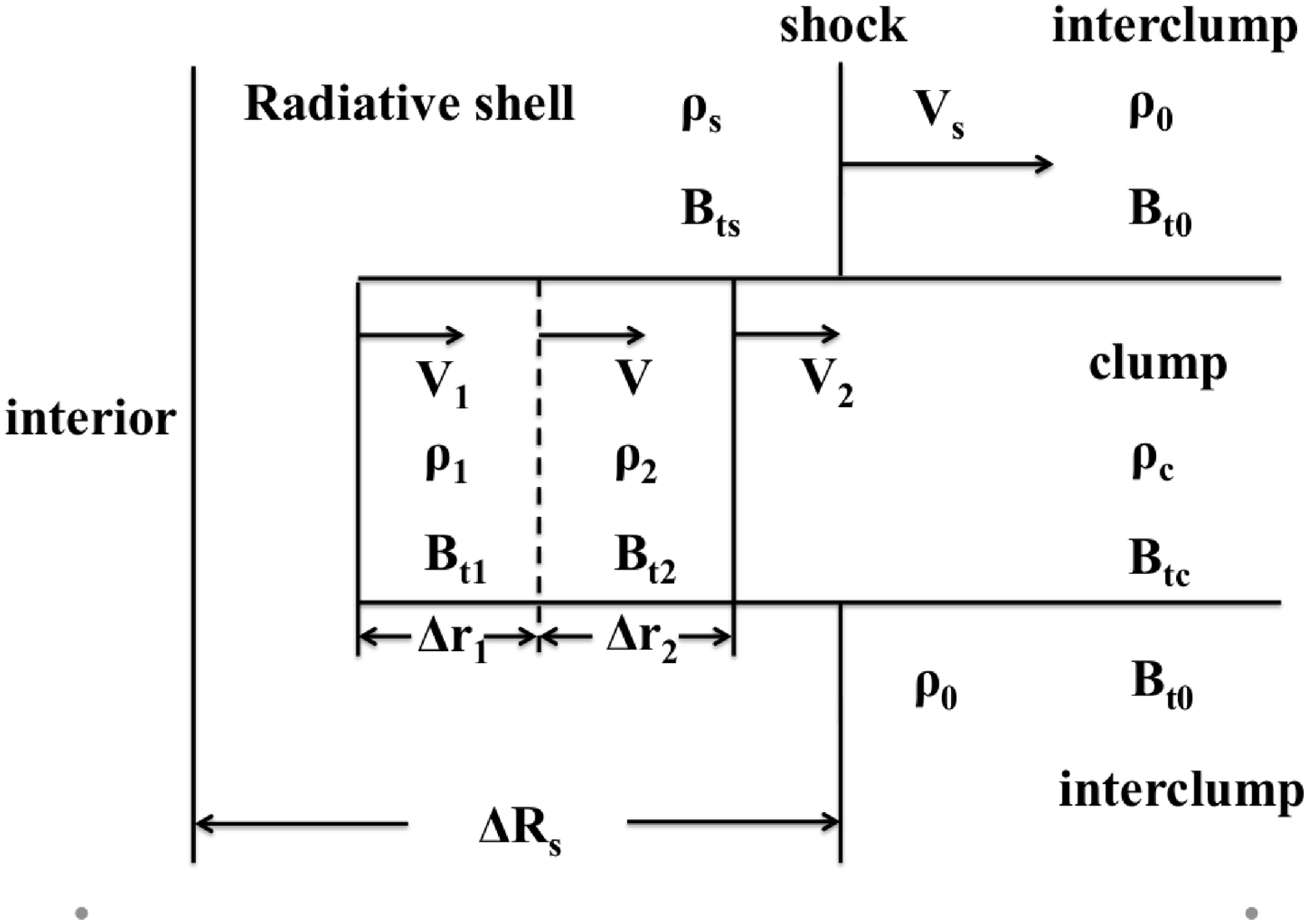}
  \caption{Schematic figure of the interaction between the radiative shell and a molecular clump.
See the text for definitions.} 
    \label{MCinteraction}
 \end{center}
 \end{figure}

We simplify the collision between the radiative shell and MC clump to a one-dimensional (1-D) problem in order to obtain an analytical solution for the structure of layers 1 and 2 (Figure \ref{MCinteraction}). We  assume both cooled layers are supported by magnetic pressure.
The thicknesses of layers 1 and 2 follow the kinetic relations
$\Delta r_1= (V-V_1)t_c$  and  $\Delta r_2=(V_2-V)t_c,$,
where $t_c$ is the  time since collision and $V_1$ and $V_2$ are the velocity for layer 1 and layer 2. $V$ is the velocity of the discontinuity between layers 1 and 2. 
Mass conservation yields
$ (V_{s}-V_1)t_c\rho_{s}=\Delta r_1 \rho_1$  and $  V_2t_c\rho_c=\Delta r_2 \rho_2 $ where $\rho_1,\rho_2$ and $\rho_c$ are density for layer 1, layer 2 and the MC clump, respectively.
Magnetic flux conservation gives
$ (V_{s}-V_1)t_cB_{ts}=\Delta r_1 B_{t1} $ and $ V_2t_cB_{tc}=\Delta r_2 B_{t2},  $
where $B_{ts},B_{t1}, B_{t2}$ and $B_{tc}$ denote the tangential magnetic field for the shell, layer 1, layer 2, and MC clump, respectively.
Magnetic pressure support for layers  1 and  2 requires
$ B^2_{t2}/8\pi=\rho_cV_2^2 $ and  $ B^2_{t1}/8\pi=\rho_{s}(V_{s}-V_1)^2, $ where we have
neglected the preshock magnetic pressure because it is only a few \% of the ram pressure.
Pressure balance at the discontinuity gives $B_{t1}^2/8\pi=B_{t2}^2/8\pi$. In the above relations, $V_s$ and $V$ are from observations while $\rho_s$ and $B_{ts}$ can be calculated for a magnetically supported shell.   At the time $t_{MC}$ that layer 1 breaks out the shell, i.e. $\Delta r_1=\Delta R_s$, all the unknown parameters can be found, using the relation between $B_{tc}$ and $\rho_c$ \citep{Crutcher12}. The  parameters for IC 443 are listed in Table \ref{IC443MC} based on \cite{dishoeck93}, \cite{Chevalier99}, and \cite{cesarsky99}.
These values are only representative, as there are expected to be variations in different parts of the remnant and among different clumps.

\section{RELATIVISTIC PARTICLE EVOLUTION AND EMISSION}

In the radiative shock models of \cite{B&C82} and \cite{Uchiyama10},
pre-existing CRs were assumed to be accelerated at the shock by DSA and then accumulated in the dense shell due to adiabatic compression.
\cite{bykov00} considered cases where there is injection of thermal particles into the acceleration process.
When clump interaction occurs \citep{Chevalier99}, CRs trapped in the radiative shell and pre-existing CRs in the clumps undergo acceleration after being swept up by the clump interaction shock and then accumulated in layers 1 and 2 respectively. We use the same number density of pre-existing CR protons 
$n_{GCR}(p)=4\pi J\beta^{1.5} p_0^{-2.76}$ as in \cite{Uchiyama10} but with a low energy cutoff of $3$ MeV and a high number density limit of 2.5 cm$^{-2}$ s$^{-1}$ sr$^{-1}$ GeV$^{-1}$ to approximate the Voyager 1 data \citep{stone13}. This result also provides a reasonable approximation up to several TeV; a detailed fit to the proton spectrum in \cite{Adriani11} does not significantly affect our results. We assume that the CR properties in clumps are the same as in the ICM and they are similar to those in the rest of the Galaxy.  {\it Fermi} observations of nearby MCs show evidence for a relativistic particle population that is similar to that observed near Earth \citep{yang13}.

We assume all the $\gamma$-ray emission comes from the shell and the clump interaction region, and we only model the radiative phase of the SNR from $t_b$ to $t_{age}$. The time evolution of the total number of CR protons $N(E,t)$ in the shell follows \citep{sturner97}
\begin{equation}
\frac{\partial N(E,t)}{\partial t}=\frac{\partial b(E,t) N(E,t)}{\partial E}+Q(E,t)-\frac{N(E,t)}{\tau_{pion}},
\label{evolution}
\end{equation}
where $b(E,t)=-dE/dt$ is the energy loss term for protons including adiabatic expansion and Coulomb collisions.  $\tau_{pion}$ characterizes the loss time of CR protons due to $p-p$ interactions, $\tau_{pion}(E_p)=1/(c\beta_p n\sigma_{pp}$), where the $p-p$ cross section $\sigma_{pp}$ is from \cite{Kelner06}.   The particle injection rate is
\begin{equation}
Q(E,t)=\frac{Q(p,t)}{v(p)}=\frac{4\pi  R_s^2(t)V_s(t) (1-w)}{v(p)s(t)}n_{in}(p,t),
\label{injection}
\end{equation}
where $v(p)$ is the proton velocity for momentum $p$, $s(t)=n_s(t)/n_0$ is the total density compression ratio in the shell, and $w$ is the surface area filling factor. 
The shell radius and velocity, $R_s(t)$ and $V_s(t)$, are from the analytical solution of \cite{Cioffi88}.
For the injected CR number density after acceleration, $n_{in}(p,t)$, we considered two cases. 
In one, DSA at the shock and further adiabatic compression in the cooling shell gave  \citep{Uchiyama10}
\begin{equation}
n_{in}(p,t)=[s(t)/\lambda_s]^{2/3}n_{DSA}([s(t)/\lambda_s]^{-1/3}p),
\end{equation}
where $n_{DSA}(p)$ is the number density of pre-existing CRs after DSA and $\lambda_s=4$ is the strong shock compression ratio. The other case  is pure adiabatic compression, with $n_{in}(p,t)=s^{2/3}(t)$ $n_{GCR}[s^{-1/3}(t)p]$. 

After obtaining $N(E,t_{age})$ in the shell from equation (\ref{evolution}), the total number of CR protons in layer 1, $N_1(E,t_c)$, and layer 2, $N_2(E,t_c)$, could be found in the same way as for the clump interaction from $t_{age}$ to $t_{age}+t_{MC}$, but with a different source term. Because the collision time $t_c$ is coupled to the filling factor $w$ in the 1-D case,  we set $t_c=t_{MC}$ and then fit the data by varying $w$; this gives the minimum value for $w$. Our model only applies to the situation $t_c\le t_{MC}$; if $t_c>t_{MC}$, layer 1 breaks out of the shell, complicating the situation,
and the emission is expected to fade.   With the parameters of interest here, we find the breakout time  $t_{MC}=\Delta R_{s}/(V_{s}-V_1) \sim  0.4 {\rm~kyr}\ll t_{age}-t_b$, so we can neglect the evolution of the radiative shell when we calculate the structure of layers 1 and  2.

The shock at layer 2 is a slow nonionizing shock so we only consider the pure adiabatic case,
\begin{equation}
Q_{2}(E,t_c)=\frac{Q_{2}(p,t_c)}{v(p)}=\frac{4\pi R^2_{s}(V_2-V)w}{v(p)}s_2^{2/3}n_{GCR}(s_2^{-1/3}p).
\end{equation} 
For layer 1, 
\begin{equation}
Q_{1}(E,t_c)=\frac{Q_{1}(p,t_c)}{v(p)}=\frac{4\pi R^2_{s}(V-V_1)w}{v(p)}n_{in,1}(p).
\end{equation} 
In the pure adiabatic case, $n_{in,1}(p)=s_1^{2/3}$ $n_{age}(s_1^{-1/3}p)$, where $n_{age}(p)$ is the number density in the radiative shell at $t_{age}$, while for the DSA case $n_{in,1}(p)=(s_1/\lambda_1)^{2/3}$ $n_{DSA,1}((s_1/ \lambda_1)^{-1/3}p)$ where $n_{DSA,1}(p)$ is the number density  $n_{age}(p)$ after DSA.   $s_1$ and $s_2$ are the density compression ratios for layers 1 and 2. $\lambda_1$ is the shock compression ratio for layer 1 and is calculated from the jump conditions for a perpendicular shock \citep{draine93}.  
Equation (\ref{evolution}) is calculated numerically with the Crank-Nicolson method for both shell and  clump interaction region. 

The efficiency of DSA is limited by the available particle acceleration time, $t_{acc}$.  
For shell evolution $t_{acc}=t_{age}-t$, while for clump interaction $t_{acc}=t_{MC}-t_c$.
By comparing $t_{acc}$ with the timescale for DSA $t_{DSA}\simeq (10/3)\eta c r_g v_{shock}^{-2}$, where $r_g$ is the gyro radius and $\eta\ge 1$ is the gyro factor, we introduce an exponential cutoff at $p_{max}$ \citep{Uchiyama10} for the CR number density in both the shell and layer 1. Here, we assume $\eta=1$ to obtain the most efficient DSA.
When $p>p_{max}$, $t_{acc}<t_{DSA}$, limiting the energy particles can reach.
For a typical radiative SNR, $p_{max}\approx 96 $ $ (10/\eta)(t_{acc}/20{\rm ~kyr})$ $(B_0/5~\mu {\rm G})(V_s/100{\kms})^2$ ${\rm ~GeV}/c$.
For a shock with $V_s\la 100 \kms$ running into a dense medium, the shock precursor is not strongly ionized and ion neutral damping of Alfven waves becomes important. As a result, high energy CR particles can escape the DSA site, bringing a steepening factor $p_{br}/p$ to the particle spectrum when $p>p_{br}$; the break momentum is $p_{br}\approx 9.4(B_0/1{\rm~\mu G})^2$ $(T/10^4{\rm~K})^{-0.4}$ ${\rm (1~cm}^{-3}/n_0)(1~{\rm cm}^{-3}/n_i)^{1/2}{\rm  ~GeV}/c $ \citep{Malkov11}. 
 
Trapping of CR particles in the cooling region due to the high tangential magnetic field might also limit DSA.
By comparing the column density of the downstream acceleration region $N_c=n_0V_st_{DSA}$ with that for gas to cool down to $ 10^4$ K and become radiative $ N_{cool}\approx 3\times 10^{17}(V_s/10^2\kms)^4$  for $60<V_s<150\kms$ \citep{Mckee87}, 
we obtain a critical momentum $p_{cr}\approx 9.1 (1/\eta)(B_0/1{\rm~\mu G)(1~cm}^{-3}/n_0)$ $(V_s/100 \kms)^5{\rm  ~GeV}/c $. When $p\ga p_{cr}$, particles may be trapped in the dense region before DSA is complete.

The $\gamma$-ray emission from $\pi^0$-decays in the radiative shell and clump interaction region is calculated based on \cite{Kamae06},  which gives results  consistent with \cite{Dermer86} within $15\%$.   The scaling
factor $\chi$ for helium and heavy nuclei is taken to be 1.8 \citep{mori09}. The $\gamma$-ray flux density at $E_{\gamma}$ is \citep{sturner97,Kamae06}
\begin{equation}
F_{\pi^0}(E_{\gamma},t_{age})=\frac{\chi E_{\gamma}}{4\pi d^2}\int dV\int^\infty_{E_{p,thresh}}dE_{p} 4\pi n_{s}(t_{age})J_{p}(E_{p},t_{age})\frac{d\sigma (E_{\gamma},E_p)}{dE_{\gamma}}.\\
\end{equation}
Given the proton flux density $J_{p}(E,t_{age})=c\beta_{p}n_{in}(E,t_{age})/4\pi $ and  assuming that the shell is uniform, after some calculation we obtain
\begin{equation}
F_{\pi^0}(E_{\gamma},t_{age})
=\frac{\chi E_{\gamma}cn_{s}(t_{age})}{4\pi d^2}\int^\infty_{E_{p,thresh}}dE_{p} \beta_{p}\frac{d\sigma (E_{\gamma},E_p)}{dE_{\gamma}} N(E_{p},t_{age}).
\label{emission}
\end{equation}
In the above equation, $n_{s}(t_{age})$ and $N(E_p,t_{age})$ need to be replaced by $n_{1}(t_{MC})$ and $N_1(E_p,t_{MC})$ for layer 1, and $n_{2}(t_{MC})$ and $N_2(E_p,t_{MC})$ for layer 2
when calculating emission from the clump interaction region.

We calculated the electron bremsstrahlung component for our models, assuming a 1 to 100 abundance ratio of cosmic ray electrons to protons.  The leptonic emission is not significant compared to the hadronic emission in view of the low abundance of electrons.

\section{RESULTS AND DISCUSSION}

\begin{figure}[htb]
 \begin{center}
  \includegraphics[width=\textwidth]{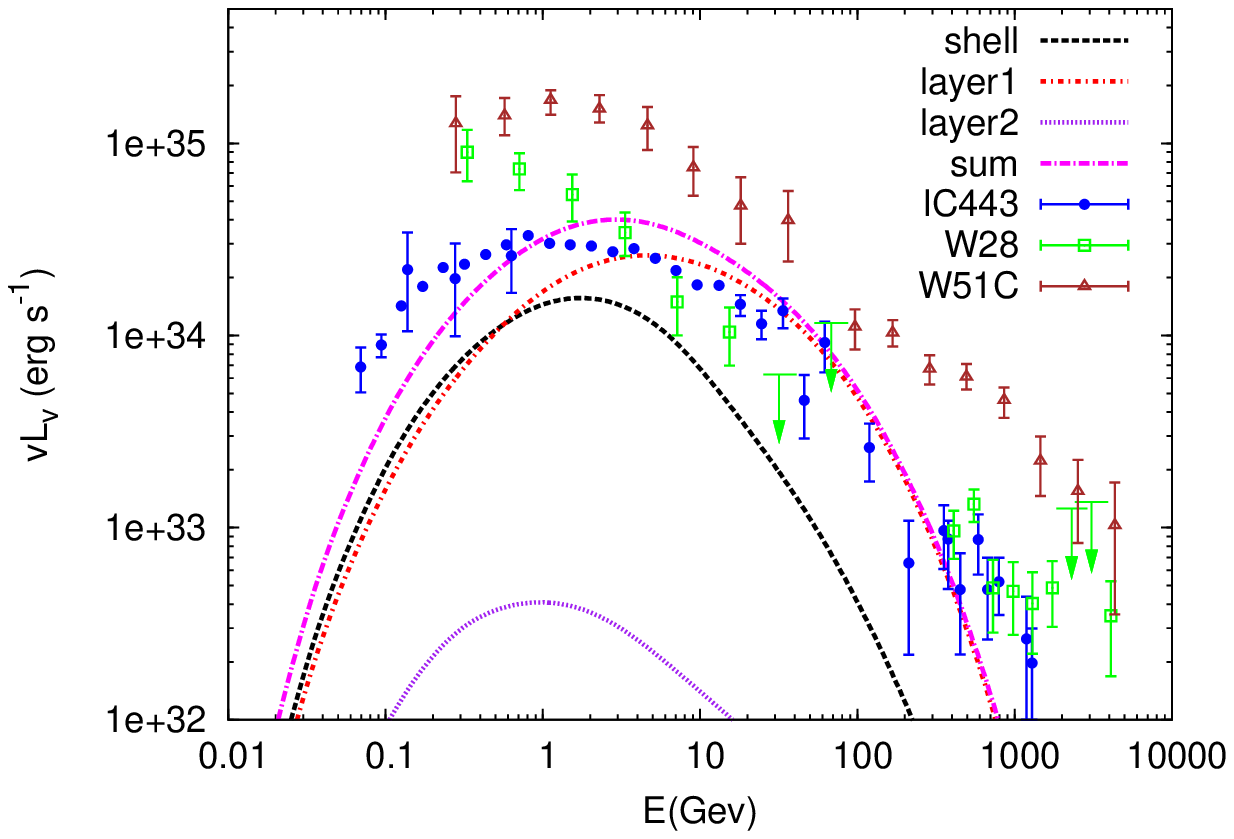}
  \caption{
$\gamma$-ray emission from IC 443 in a model with DSA plus adiabatic compression.
The data points for IC443 are from \cite{Ackermann13,Albert07,Acciari09,tavani10} with a distance of 1.5 kpc; W28 from \cite{Abdo10c} and \cite{Aharonian08} with a distance of 2 kpc; W51C from \cite{Abdo09} and \cite{Aleksic12} with a distance of 4.3 kpc \citep{Tian13}.}
    \label{IC443GR}
 \end{center}
 \end{figure}

\begin{figure}[htb]
 \begin{center}
  \includegraphics[width=\textwidth]{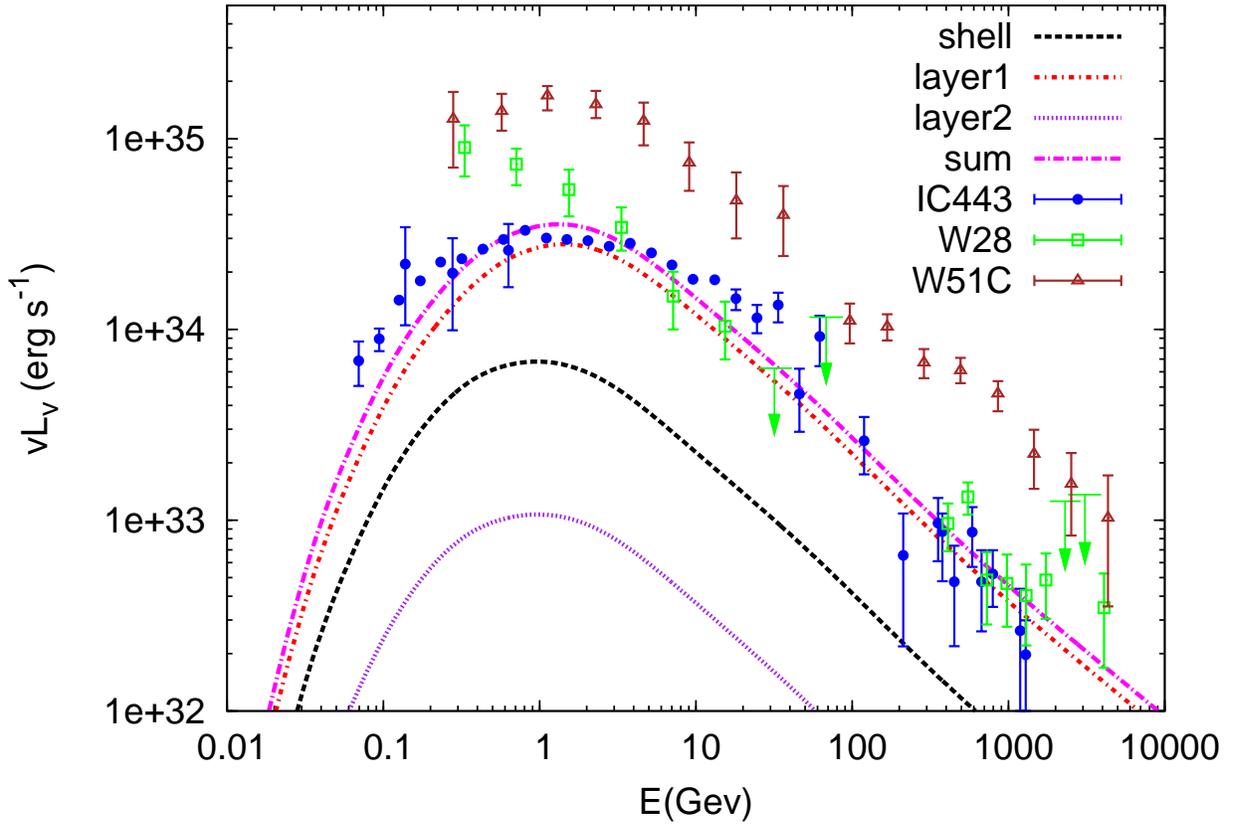}
  \caption{$\gamma$-ray data as in Fig. \ref{IC443GR} compared to the pure adiabatic compression model for IC 443. } 
    \label{IC443GR_ad}
 \end{center}
 \end{figure}

We calculated the emission from IC 443 for both standard DSA and pure adiabatic cases, for our radiative shell plus MC clump interaction model and the parameters shown in Table \ref{IC443MC}.    
Ion neutral damping and a finite acceleration time were taken into account in the DSA simulation but not  the limited cooling region argument due to its uncertainty. The ionization in the layer 1 precursor is low so we only considered ion neutral damping for the shell, finding that $p_{br}\sim \rm  10$ GeV/c gives a good fit to the GeV part of the spectrum with $w=8\%$ (Figure \ref{IC443GR}). The resulting $\gamma$-ray spectrum is narrower than the observed spectrum, falling below the observed emission at high energy due to the limited acceleration time. The observed spectra from GeV to TeV energies have slopes which are similar to the input CR spectrum. While pure adiabatic compression maintains the shape of the input CR spectrum,  pion-decay emission also traces the energy distribution of the parent spectrum above a few GeV.  We found that the pure adiabatic case can approximately fit the spectra of IC 443 from GeV to TeV energies, but with a higher $w\approx 21\%$ (Figure \ref{IC443GR_ad}), which implies that the remnant is in a special phase of evolution. Alternatively, a higher value for the ICM density would reduce the value of $w$.

The $\gamma$-ray spectrum of  W28 has a similar shape to that of  IC 443, except for the low energy part \citep[Figure \ref{IC443GR};][]{Abdo10c,Aharonian08}.
W51C's $\gamma$-ray spectrum also has a shape similar to IC 443 but the luminosity is larger \citep[Figure \ref{IC443GR};][]{Abdo09,Aleksic12}. 
W51C is a large remnant compared to the other two, and may have an unusually large
energy \citep{koo97,koo05}, which could account for the high luminosity.
We do not attempt detailed models for W28 and W51C, but note the similar spectral shapes for the three remnants may be a result of the similar parent CR spectrum.

Other possible tests of the models are the shocked MC mass and the relative intensities of the three components: radiative shell, layer 1 and layer 2. In the pure adiabatic case the shocked MC mass required for IC 443  is $m_{MC}\approx V_2t_{MC}\rho_c 4\pi R^2_s w\approx 500 \Msun$ which is more consistent with the molecular observations \citep{dickman92,Lee08}  than the $190 \Msun$ in our DSA model.   In the pure adiabatic case, the $\gamma$-ray emission from IC 443 is naturally dominated by the MC interaction region, especially layer 1, while for the DSA case the emission from the MC interaction region is comparable to the shell component at low energy but becomes dominant at high energy.   However, the surface brightness is coupled with the angle between the collision direction and the line of sight.
Considering the uncertainty in both theoretical models and observations, these tests are not definitive.

Here we have sought a model for the $\gamma$-ray emission from SNRs that is consistent with the emission properties given in Section 2.
The correlation of GeV and TeV emission with molecular clump interaction implies that the $\gamma$-ray emission is related to slow radiative shock waves in dense matter. 
Standard DSA is not efficient at high energies. Pure adiabatic compression could reproduce the ratio of TeV to GeV emission but requires a large covering factor (Figure 3). Particle acceleration process in the middle aged SNRs are still not very clear.  There are other possibilities; \cite{bykov08} have found nonthermal X-ray emission near a clump interaction region in IC 443 which they interpret as the result of ejecta knots hitting the molecular gas. More detailed observations of the remnants discussed here, as well as other remnants with molecular cloud interaction, would improve our understanding of CR acceleration in SNRs.

\acknowledgements
We are grateful to Charles Dermer for the code used to calculate the $\pi^0$-decay emission, and to A. Bykov and the referee for useful comments.
This research was supported in part by NASA {\it Fermi} grant NNX12AE51G.

\clearpage

\begin{table}[ht]
\centering
\caption{Basic parameters for the IC 443 model}
\begin{tabular}{lr}
\hline\hline
SNR dynamics&\\
\hline
Explosion energy, $E$ & $0.45 \times 10^{51}$ erg\\
Age, $t_{age}$ &22.3 kyr\\
SNR radius, $R$ &7.4 pc\\
Shock velocity, $V_{s}$ &100 km/s\\
Shock compression ratio, $\lambda_s$ &4\\
Simulation start time,  $t_b$ &4.4 kyr\\
\hline\hline
MC clump and ICM &\\
\hline
Preshock ICM density, $n_0$ & 15 cm$^{-3}$\\
Magnetic field in ICM, $B_0$ &5 $\mu$G\\
MC clump density, $n_c$ &$1\times 10^4$ cm$^{-3}$\\
Magnetic field in MC clump, $B_c$ &51 $\mu$G \\
\hline\hline
Radiative shell and MC clump interaction region &\\
\hline
Discontinuity velocity of clump shock, $V$  &25 km/s\\
MC clump interaction break out time, $t_{\rm MC}$  &0.37 kyr\\
Density in the radiative shell at $t_{age}$,  $n_s$ & $9\times 10^2$ cm$^{-3}$\\
Magnetic field in the radiative shell at $t_{age}$,  $B_{ts}$ &$3\times 10^2$ $\mu$G\\
Density in layer 1,  $n_1$ &$6\times 10^3$ cm$^{-3}$\\
Magnetic field in layer 1,  $B_{t1}$ & $2\times 10^3$ $\mu$G\\
Layer 1 velocity,  $V_1$ &12 km/s\\
Shock compression ratio for layer 1, $\lambda_1$ &3.4\\
Density in layer 2,  $n_2$ &$5 \times 10^5$ cm$^{-3}$\\
Magnetic field in layer 2,  $B_{t2}$ &$2\times 10^3$ $\mu$G\\
Layer 2 velocity,  $V_2$ &26 km/s\\
\hline
\end{tabular} 
\label{IC443MC}
\end{table}

\end{document}